\documentclass{emulateapj}

\newcommand{\ctt}{ ${\cal C}_\ell^{TT}$ }
\newcommand{\cte}{ ${\cal C}_\ell^{TE}$ }
\newcommand{\cee}{ ${\cal C}_\ell^{EE}$ }



\long\def\comment#1{}

\def\etal{{\it et al.~}}

\def\W2{{\cal W}}

\def\be{\begin{equation}}
\def\ee{\end{equation}}
\def\bea{\begin{eqnarray}}
\def\eea{\end{eqnarray}}

\def\C{{\cal C}}
\def\l{{\ell}}

\def\cmm2{{\,\rm cm^{-2}}}
\def\cm2{{\,{\rm cm}^2}}
\def\cmm3{{\,{\rm cm}^{-3}}}
\def\gcmm3{{\,{\rm g\,cm^{-3}}}}

\def\fun#1#2{\lower3.6pt\vbox{\baselineskip0pt\lineskip.9pt
  \ialign{$\mathsurround=0pt#1\hfil##\hfil$\crcr#2\crcr\sim\crcr}}}

\def \ie {{\it i.e. }}

\newcommand{\MNRAS}{Mon. Not. R. Astron. Soc.}

\usepackage{epsf}

\begin{document}


\submitted{ApJ in press}

\title{The CMB Quadrupole in a Polarized Light}

\author{Olivier Dor\'e$^{1}$, Gilbert P. Holder$^{2}$, \& Abraham
Loeb$^{2,3}$} \affil{ {}$^{1}$Department of Astrophysical Sciences,
Princeton University, Princeton NJ 08544\\
{}$^{2}$School of Natural Sciences, Institute for Advanced Study, 
Princeton, NJ 08540\\
{}$^{3}$Astronomy Department, Harvard University, Cambridge, MA 02138
}

\begin{abstract}
The low quadrupole of the cosmic microwave background (CMB), measured by
COBE and confirmed by WMAP, has generated much discussion recently. We
point out that the well-known correlation between temperature and
polarization anisotropies of the CMB further constrains the low multipole
anisotropy data.  This correlation originates from the fact that the
low-multipole polarization signal is sourced by the CMB quadrupole as seen
by free electrons during the relatively recent cosmic history.
Consequently, the large-angle temperature anisotropy data make restrictive
predictions for the large-angle polarization anisotropy, which depend
primarily on the optical depth for electron scattering after cosmological
recombination, $\tau$. We show that if current cosmological models for the
generation of large angle anisotropy are correct and the COBE/WMAP data are
not significantly contaminated by non-CMB signals, then the observed \cte
amplitude on the largest scales is discrepant at the $\sim 99.8\%$ level
with the observed \ctt for the concordance $\Lambda$CDM model with
$\tau=0.10$.  Using $\tau=0.17$, the preferred WMAP model-independent value,
the discrepancy is at the level of 98.5\%.
\end{abstract} 


\maketitle

\section{Introduction}

The low quadrupole (and first few multipoles) of the cosmic microwave
background (CMB), measured by COBE \citep{Be96} and confirmed by WMAP
\citep{Be03}, has generated much discussion recently, with several papers
offering various possible causes for suppressed power on very large scales
(e.g., Bond 1995, Efstathiou 2003a, Bridle \etal 2003, Contaldi \etal 2003, Tegmark \etal 2003, Cline
\etal 2003, Feng \& Zhang 2003, among others).  
\nocite{Ef03a,Con03,Te03,cline03,Fe03} Often in
the case of measurements suggestive of new physics, the obvious way to 
advance is to
perform a better experiment.  However, the current measurements of the
temperature anisotropy power spectrum of the CMB on large angular scales
are already limited only by how well the Galaxy can be removed
\citep{Be03}, and so the prospects for improved measurements are 
poor.  In
this work we point out that polarization measurements on large angular
scales can test whether the temperature anisotropy on these scales is
indeed generated within the standard cosmological framework.

The correlation between temperature and polarization anisotropies is
well-known \citep{ZaSe97}. 
It has been detected by \citet{Le02} on intermediate angular scales
and measured by the WMAP experiment on the scales of interest here
\citep{Ko03}. This correlation has been used to predict the polarization
pattern on the sky from the observed temperature pattern \citep{jaffe03}.
The correlation between temperature and polarization arises because the
source of the polarization is Thomson scattering of the quadrupole
anisotropy in the temperature of the radiation field. Spatial fluctuations
in the monopole and dipole of the temperature field at the time of
recombination seed higher multipole anisotropies by free-streaming (for a
recent review see Hu \& Dodelson 2002). Most of the large scale polarized
signal we consider here originated from Thomson scattering of the CMB
quadrupole at the relatively recent epoch following the reionization of the
Universe \citep{Za97}.  \nocite{hu02} The quadrupole temperature anisotropy
seen by free electrons at this epoch receives a significant contribution
to its $k$-space kernel from density fluctuation modes on scales $k^{-1}$
that also contribute to the present-day quadrupole \citep{TeZa02}.

For a given realization of the CMB sky, the measured temperature anisotropy
power at a given multipole, \ctt will have some amount of intrinsic
scatter, as will the polarization anisotropy power, \cee and
the cross power spectrum, \cte. Importantly, these measures of
the power are correlated. If \ctt is measured to be low, then one would
expect a low measure of \cte and also a low value of \cee.  Therefore,
if measures of \cte and \cee at low multipoles (large angular 
scales) are not ``anomalously'' low, then this would exacerbate the 
current tension with theoretical models \citep{Sp03}. 

$\Lambda$CDM is currently the standard cosmological model, and so we choose
to assume the best-fit WMAP cosmological parameters and study how likely it
is, from a frequentist perspective, that the observed data are simply
realizations of this model.  If the assumed model is in fact correct, then
the correlations between observables provide statistical consistency
checks. For example, flukes of cosmic variance should be partly correlated
between temperature and polarization observables, so it might be expected
that large outliers in the temperature data will have counterparts in the
polarization data. Here, we present results of consistency tests using
current data, and estimate the range of possible future data that could be
comfortably accommodated within the currently accepted cosmological models.

\section{TT--TE correlations}

\begin{figure*}[t]
\epsscale{1.4}
\plotone{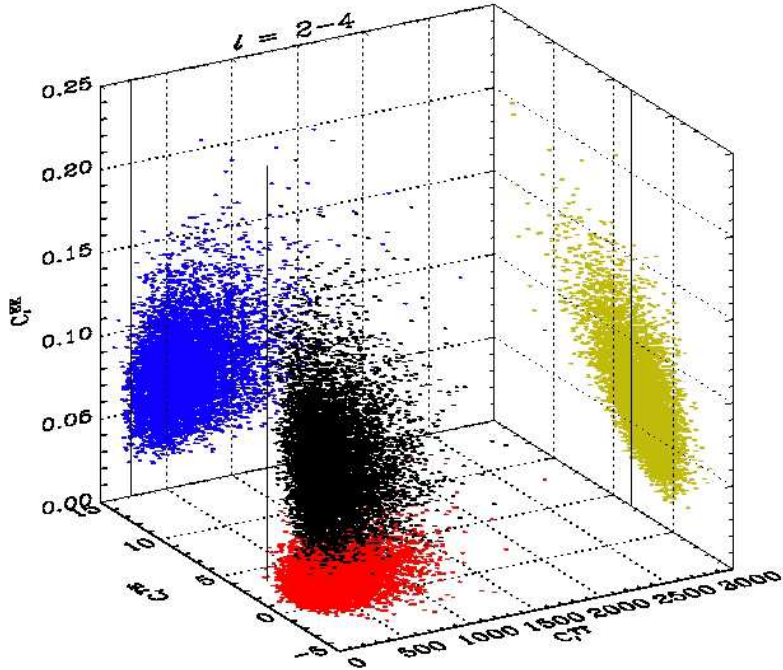}
\caption{ 3D view (black points) of realizations of WMAP best fit model for
$\ell=2-4$, assuming a noise model comparable to one year of WMAP
observations.  Projections are shown on the sides of the ``box.''  Center
dark vertical line shows measured WMAP point in this space; there is no
\cee data released at this time so no constraint along the vertical
axis. The horizontal (\cte,\ctt) plane corresponds to Figure
\ref{fig:wmap1yr_l2} whereas the vertical left (\cee,\ctt) plane
corresponds to Figure \ref{fig:wmap1yr_ee}. 
} 
\label{fig:wmap1yr_3D}
\end{figure*}
The correlations between the temperature and polarization power spectra can
be used as a powerful consistency test. For example, it is possible to
use the measured \cte to construct a probability distribution for \ctt
and compare this to measured values of \ctt. Alternatively, the measured
\ctt can be used to estimate the most likely values of measured \cte or
to forecast future \cte or \cee measurements. We present examples of
both these calculations below.

Explicit expressions for the covariance between power spectra, 
assuming Gaussian uncorrelated noise and Gaussian beams, are given
by \citet{ZaSpSe97} and are useful for understanding the nature of
the covariances.
In all that follows we neglect beam effects since we are interested in 
very large scales, where $\ell$ is much smaller than
the inverse of the beam size (in radians).  All $\C_\l^{X}$ (where
$X$ corresponds to $TT$,$TE$, or $EE$) are in units of $\mu {\rm K}^2$.

\section{Monte Carlo Methodology}

We generate $10^5$ realizations of the WMAP-only best-fit cosmological
model $a_{\l m}$ ($\Omega_mh^2=0.13$, $\Omega_bh^2=0.023$, $h=0.68$,
$n_s=0.97$, $A=0.8$ and $\tau = 0.10$ as in table 1 of \citet{Sp03}) and
for $\ell <40$ we construct theoretical joint distributions of \cee, \cte
and \ctt for this model. The scatter in these quantities around the input
model is both due to cosmic variance and noise, and the scatter between
different power spectra is somewhat correlated. We use the 1 year WMAP data for
\ctt and \cte from the corresponding  publicly--available ASCII files 
\footnote{\texttt{http://lambda.gsfc.nasa.gov}}.

We use a simple noise model which is as close as possible to that reported
by the WMAP experiment after one year of operation. Specifically we assume
that the white noise level corresponds to the use of 16 W-Band, 8 V-band
and 8 Q-band channels (see table 1 in \citet{Be03}) with an effective sky
fraction of $0.86/1.14$ \citep{Ve03}, so that we get $w_{T}^{-1/2}=
1.07\times 10^{-1} \mu$K per (0.21)$^2$ sq. deg. pixel with a corresponding
$w_{E}=w_{T}/2$.  We add this white noise directly to the generated $a_{\l
m}$. This simple simulation scheme neglects the weak extra power at low
$\l$ originating from the residual $1/f$ noise \citep{Hi03}.  We neglect
any non-gaussian systematic uncertainty that may be caused by improper
subtraction of the galactic component \citep{Be03,Ef03b}. However, to
assess the significance of our assumed level of noise, we also consider
simulations where the noise amplitude, $w_{T}^{-1/2}$, was arbitrarily
scaled up by a factor of $\sqrt{2}$, so that the noise contribution to the
$\C_\l^X$ errors is doubled. This excess noise could be either some unknown
component of low frequency noise or residual galactic contamination
(although the latter should probably not be modeled as a white noise).

Furthermore, to reduce the effects of covariance due to the real cut sky we
consider bins of width $\Delta \ell=2$--$4$, where $\C_\l^X$s are
assumed constant. With such binned power spectrum estimates, the
covariance between neighboring bins should be very small and our Monte
Carlo results should approximate reasonably well the actual covariance
of the properly measured $\C_\l^X$ (we estimate that the correlation
between our lower bins is less than 10\%). It is important to note that our
procedure is only approximate and it is therefore not strictly
appropriate to compare our results directly to WMAP data; however we
do not expect that more accurate simulations would alter significantly
our conclusions. 

\section{Monte Carlo Results}

\begin{figure}[tb]
\centerline{
\epsscale{1.3}
\includegraphics[angle=90,width=0.5\textwidth]{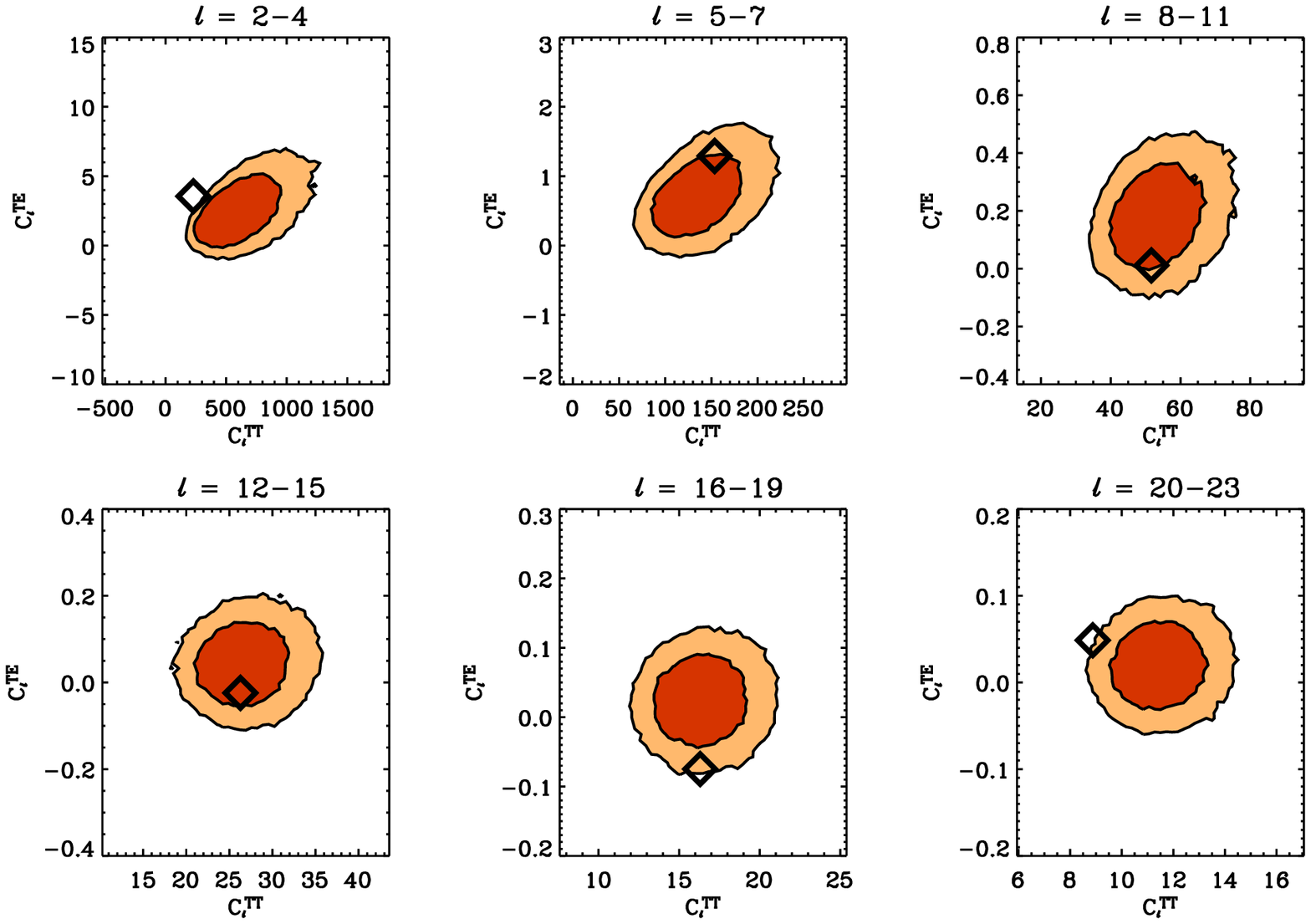}}
\centerline{
\includegraphics[angle=90,width=0.5\textwidth]{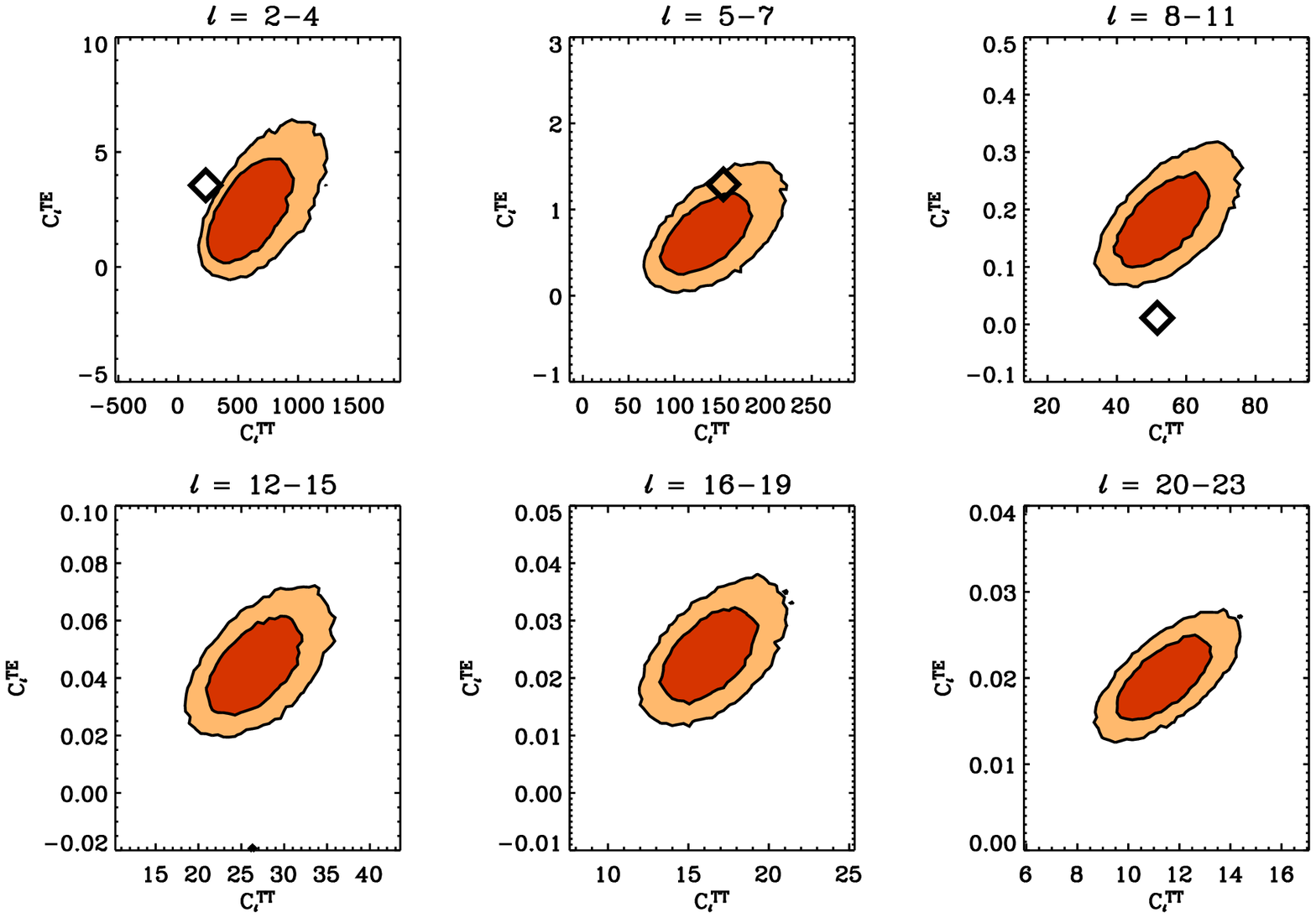}}
\caption{Realizations of \cte and \ctt for noise levels typical of one year
of WMAP observations (2 upper rows) and for cosmic variance only (2 lower
rows). WMAP data points are marked and the measurement noise and cosmic
variance are included in the realizations (\ie, it is appropriate for the
points to not have error bars).  The data points in the 2 lower 
rows are shown only for reference
(as they contain noise).}  
\epsscale{1.0}
\label{fig:wmap1yr_all}
\end{figure}
\begin{figure}[b]
\centerline{
\includegraphics[angle=0,width=0.5\textwidth]{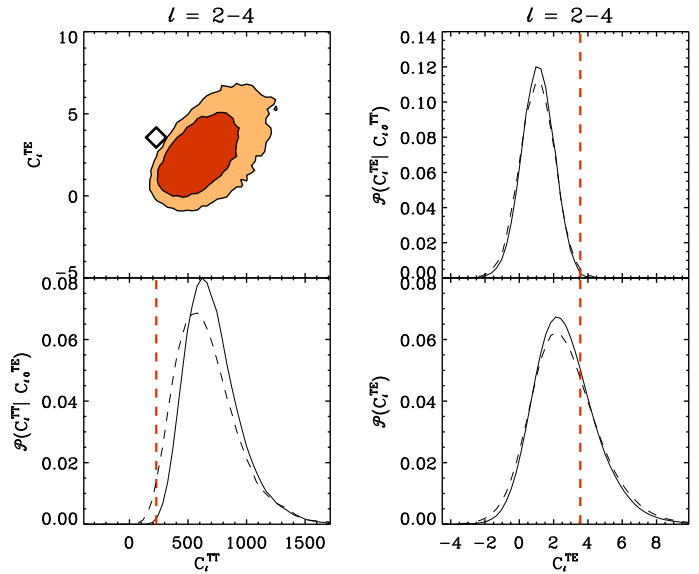}}
\caption{Lowest multipole bin distribution of \cte and \ctt assuming the
WMAP best fit cosmological model and one year of observation (top left).
The bottom right panel shows the likelihood distribution of \cte with no
information on \ctt, and the top right panel shows the likelihood subject
to the constraint that \ctt has the measured value. The bottom left panel
shows the likelihood of the \ctt given the observed \cte. Dashed vertical
lines indicate the measured WMAP values. The dashed distributions
correspond to simulations where the noise level has been arbitrarily
increased by a factor $\sqrt{2}$.}
\label{fig:wmap1yr_l2}
\end{figure}

The result of these realizations is an ensemble of points in 
(\ctt,\cte,\cee) space, shown in Figure \ref{fig:wmap1yr_3D}. 
Slices through this volume then provide a frequentist estimate of the
expected joint distribution of spectra for this particular model. In
particular, slices passing through observed points allow 
investigation of the conditional likelihoods in the other directions.
For example, the vertical line shows the observed WMAP data and
a histogram of points along this line would give the conditional 
probability of \cee given the observed \cte and \ctt. 

In Figure \ref{fig:wmap1yr_all} we show the distribution of 
expected \ctt and \cte that would be observed, 
including our approximate WMAP noise model.  Current
data already provide a consistency check, even with the relatively large
noise contribution to \cte. On the largest angular scales the measurements
are approaching the cosmic-variance limit for bins of size $\Delta \ell =
3$, even with only one year of WMAP data. However, on slightly smaller 
angular scales the data contain a significant noise component, as can be 
seen in Figure 8 of \citet{Ko03}.  This noise will be uncorrelated between 
the \cte and \ctt power spectra and leads to the bins with $\ell \gtrsim 8$ 
showing little correlation in Figure 2. 

\begin{figure}[t]
\centerline{
\includegraphics[angle=0,width=0.5\textwidth]{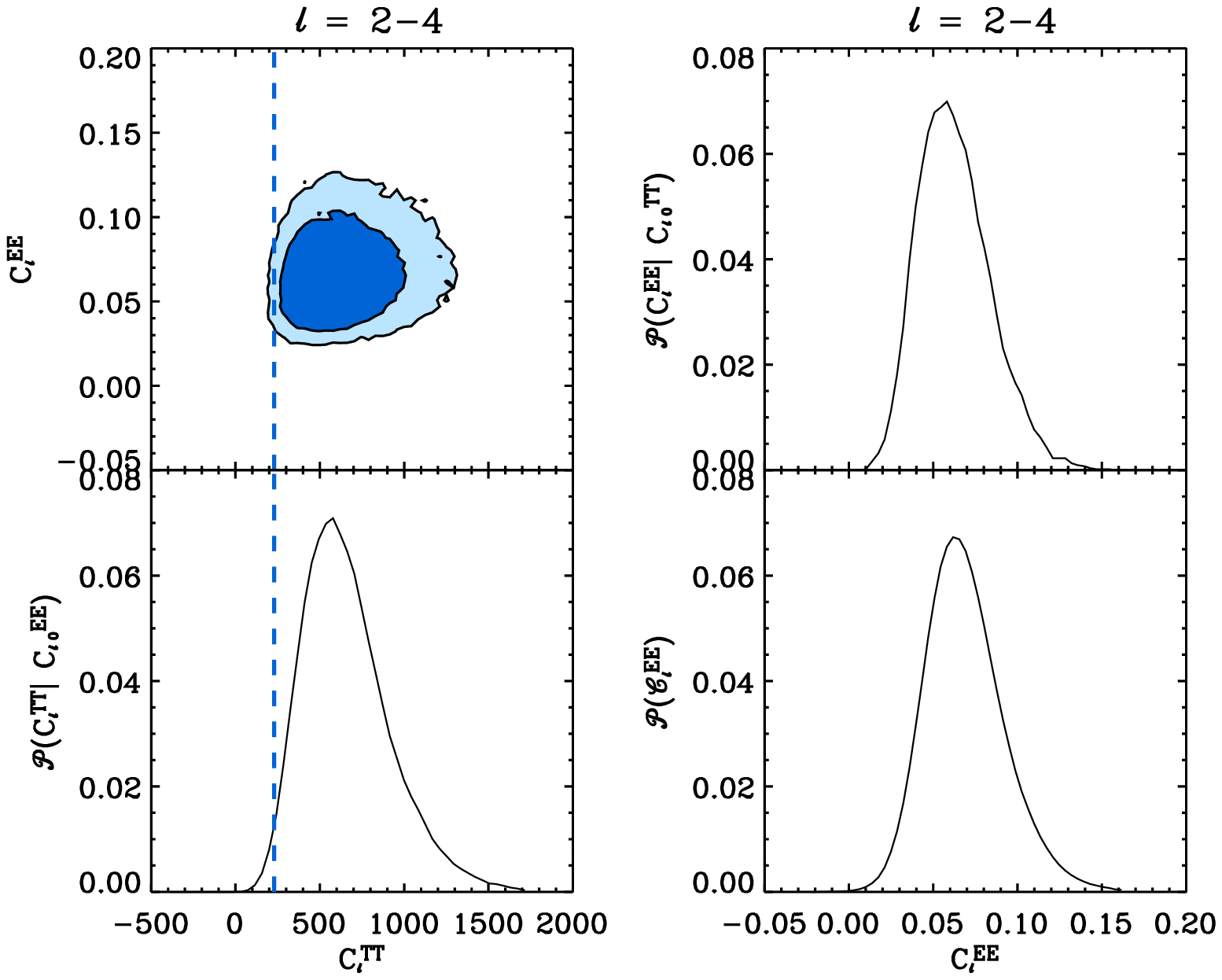}}
\centerline{
\includegraphics[angle=0,width=0.5\textwidth]{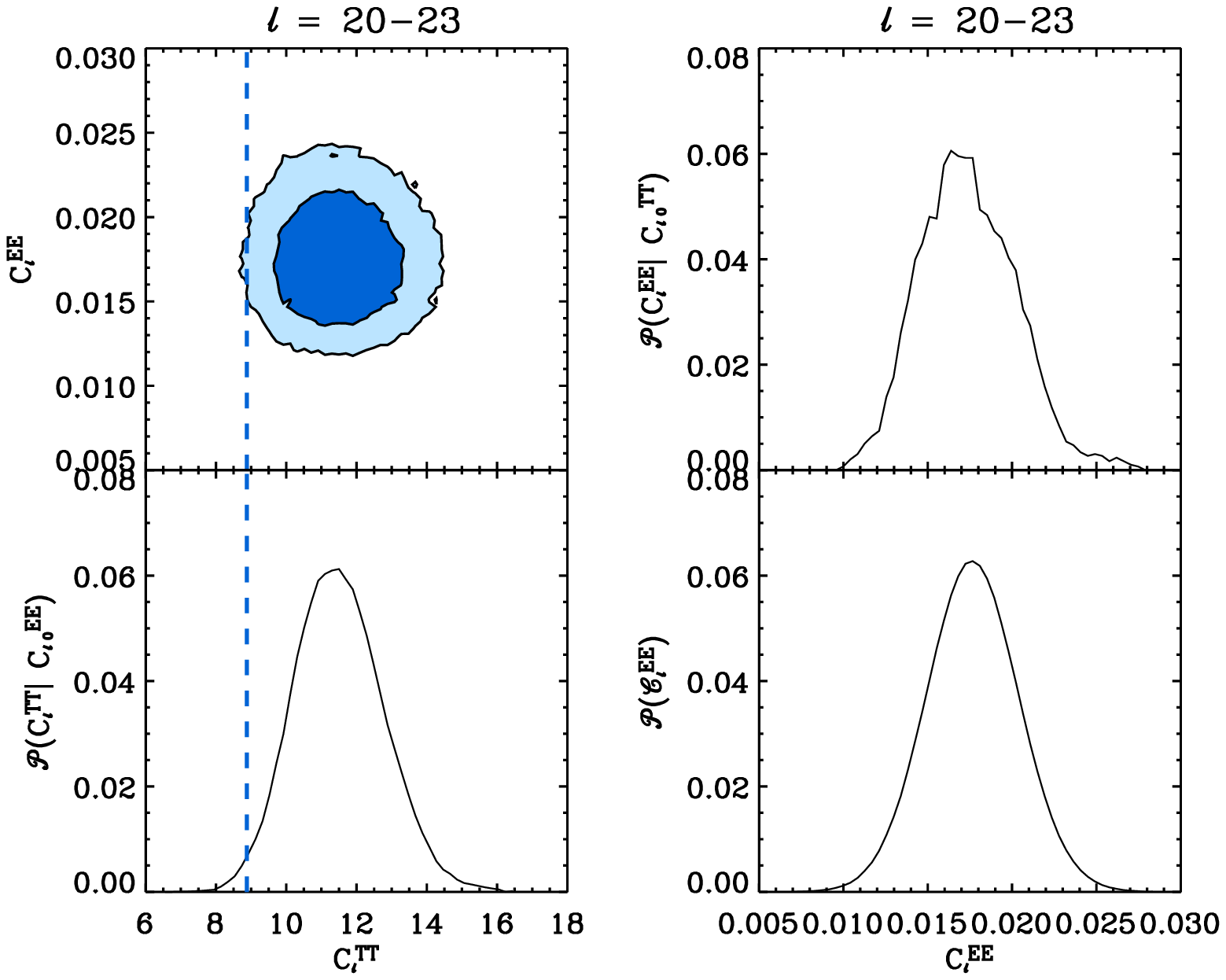}}
\caption{ Same as Figure \ref{fig:wmap1yr_l2}, now for \cee-\ctt
covariance, again assuming a noise model comparable to one year of WMAP
data. Here we show the multipole bins $2 \le \ell \le 4$ (left) and $20\le
\ell \le 23$ (right), the two largest outliers in the \ctt bins considered.
  }
\label{fig:wmap1yr_ee}
\end{figure}

In Figure \ref{fig:wmap1yr_all} we also show the \ctt-\cte correlation for
perfect measurements (no noise).  The current low-$\ell$ \ctt data are
significantly discrepant with the best-fit model, but equally remarkable is
that the \cte measurements on the largest scales are not particularly
low. The measured \cte on these scales is close to the middle of the
expected range if one ignores the correlation with the \ctt data on these
scales.  However, the probability of the measured \cte given the observed
\ctt is extremely low if this is indeed a realization of the best fit
model.  The observed \cte is in fact anomalous by not being low. This can
be seen comparing the bottom right and top right panels of Figure
\ref{fig:wmap1yr_l2}.  In the bottom right panel it can be seen that the
observed \cte on this scale is just in the range predicted by the best-fit
model (26\% of the models lie above the observed value). In the top right
panel we see that this agreement disappears when we apply the condition
that the observed \ctt is low (0.17\% of the models lie above the observed
value (or 0.85\% when the $C_\ell^X$ noise is scaled up arbitrarily by a
factor of 2). In realizations of the best-fit model it is rare that \ctt is
as low as the value measured by WMAP, but in the few realizations where
\ctt was low it was usually the case that the \cte was also
low. Correspondingly, middle-of-the-road \cte values are unlikely to appear
with low \ctt, as shown in the lower left panel.  Specifically, 99.9\% of
the \ctt values are greater than the measured one given the measured value
of \cte, while when the measured \cte is not included the fraction drops to
99.0\% (with the corresponding numbers being 99.8\% and 99.1\% when the
$C_\ell^X$ noise is arbitrarily doubled). These results are dictated by the
profile of the joint (\ctt,\cte) distribution in the top left panel of
Figure \ref{fig:wmap1yr_l2}.

Upcoming measurements of the \cee power spectrum on large scales could shed
some light on this problem. At the noise levels expected in the near future
there is little correlation between the \cee and \ctt power spectra, as
shown in Figure \ref{fig:wmap1yr_ee}, where the histograms in the right top and
bottom panels are nearly unaffected when the temperature information is
included.  The 3D plot in Figure
\ref{fig:wmap1yr_3D} again shows that the observed \cte-\ctt large angle
pair (shown as a vertical line) is exceedingly unlikely, but also shows
that there is a fairly tight correlation between \cte and \cee on these
scales. This provides an important consistency check on our understanding
of the CMB. The observed \cte-\ctt pair on these scales reduces
significantly the expected range of \cee (assuming the best fit model).

\section{Cosmological Implications}

The correlations between power spectra have several cosmological
implications.

\begin{figure}[tb]
\centerline{
\includegraphics[angle=90,width=0.5\textwidth]{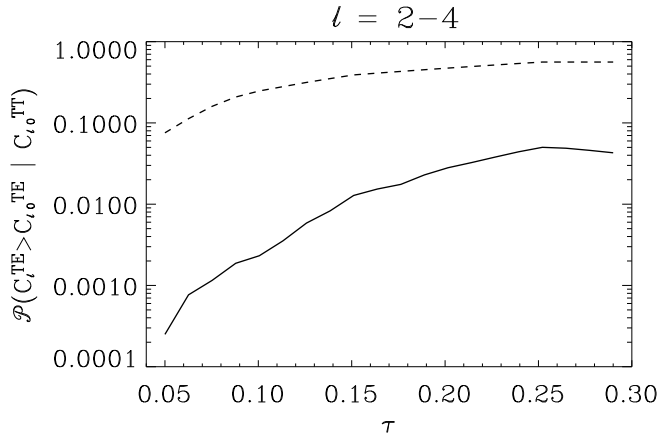}}
\caption{Probability for \cte larger than the observed value, given the
observed \ctt value for $2<l<4$, ${\cal P}(\C_\l^{TE}>\C_{\l 0}^{TE}|\C_{\l
0}^{TT})$, as a function of the optical depth value $\tau$ (solid line).
Also shown, for reference, is the same probability without the condition on
the observed \ctt value (dashed line).  All cosmological parameters are
kept the same as in our fiducial $\Lambda$CDM model except for $\tau$ and
$n_s$ that are changed simultaneously along the $\tau$--$n_s$ degeneracy
line favored by WMAP. The power-spectrum amplitude is marginalized over.
Note that ${\cal P}(\C_\l^{TE}>\C_{\l 0}^{TE}|\C_{\l 0}^{TT})$ does not
exceed 5$\%$.}
\label{fig:wmap_hightau}
\end{figure}

So far we assumed the best-fit WMAP value for the optical depth to electron
scattering after cosmological recombination, $\tau=0.10$
\citep{Sp03,Ko03} and a power-law primordial power spectrum.

Adopting a different optical depth $\tau$ has a significant but still
limited effect on our results. We repeated the analysis by considering
various optical depth values and the results are shown in Figure
\ref{fig:wmap_hightau}.  The solid line shows the probability for
2$\le\l\le$4 of having \cte larger than the observed value given the
observed \ctt value, ${\cal P}(\C_\l^{TE}>\C_{\l 0}^{TE}|\C_{\l 0}^{TT})$,
and the dashed line shows the same without the condition on the observed
\ctt value, ${\cal P}(\C_\l^{TE}>\C_{\l 0}^{TE})$. All cosmological
parameters were held fixed except for $\tau$ and $n_s$ which
were varied in
the ranges 0.05--0.29 and 0.95--1.05 respectively, along the $n_s-\tau$
degeneracy line favored by WMAP [see Figure 5 of Spergel \etal (2003)]. The
power spectrum amplitude $A_s$ was marginalized over. Other parameters
will have little impact on the large-angle polarization, as shown in
\citet{Ka03}.  A complete frequentist treatment, allowing
all parameters to vary from their best fit, is unlikely to lead to
qualitatively different results.

Increasing $\tau$
makes the observed \cte more likely, but is not sufficient to alleviate the
tension between the low \ctt and the average \cte: note that ${\cal
P}(\C_\l^{TE}>\C_{\l 0}^{TE}|\C_{\l 0}^{TT})$ does not exceed 5\%.  Note
that this tension is not included in current estimates of the optical
depth. In previous analyses of WMAP data, the \ctt likelihood is multiplied
by the bottom right panels rather than the top right panels of Figure 3.
For all but the largest angular scales (the first few multipoles) this
effect is negligible, but it is clear from Figure \ref{fig:wmap_hightau}
that the direction of the bias is such that the current estimates of $\tau$
are likely to be low.

\begin{figure}[tb]
\centerline{\includegraphics[angle=0,width=0.5\textwidth]{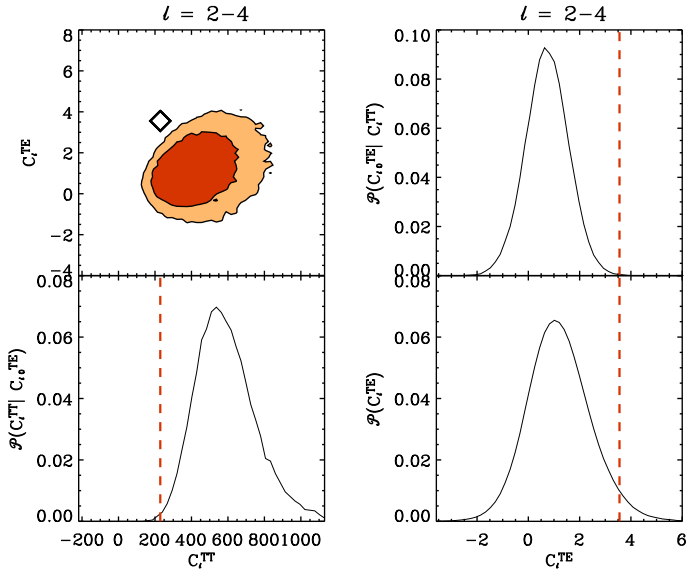}}
\caption{ Same as Figure \ref{fig:wmap1yr_l2}, except assuming cosmic
variance limited measurements and a complete suppression of the
primordial power spectrum at wavenumbers $k<k_\star = 3\times 10^{-4}\ {\rm
Mpc}^{-1}$ \citep{cline03}.  }
\label{fig:wmap_kcut}
\end{figure}

Additional suppression of the matter power spectrum on large scales tends
to reduce the discrepancy of the current \ctt data, but not by a large
amount. The mapping from matter power spectrum to temperature anisotropies
is fairly broad in Fourier space \citep{TeZa02}, so that suppression of
power on the largest scales in the matter power spectrum does not lead to a
sharp suppression only on the largest angular scales in the CMB, which is
what the data seems to suggest.  At the same time, the \cte multipoles at
$\ell \lesssim 5$ will be suppressed as well \citep{Bri03,cline03}, making
it even more difficult to match the observed \cte-\ctt pair as illustrated
in Figure \ref{fig:wmap_kcut}. Whereas now only 3.2\% of the models lie
above the observed \cte value for our lowest $\ell$ bin (as compared to the
previous 26\%), less than 0.02\% (as compared to 0.17\%) lie there if we
include the measured \ctt.

\section{Conclusions}

The correlations between the temperature and polarization power spectra
provide a powerful consistency check on CMB anisotropy measurements. Cosmic
variance fluctuations should be correlated between power spectra. By
assuming the best-fit cosmological model we have shown that the observed
\cte-\ctt data at $2 \le \ell \le 4$ are alarmingly large outliers. Much
has been made recently of the degree to which the \ctt data on these scales
is anomalously low, but it is almost equally alarming that the \ctt is low
and \cte is apparently typical. In most realizations of the WMAP best-fit
model with low \ctt on large scales, \cte is also low.  It is extremely
unlikely to see \cte as high as the measured value, given the low observed
\ctt. This correlation is not currently included in likelihood analyses of
CMB data, but should be relatively easy to incorporate in future
work. Current estimates of the optical depth are likely biased
low. Prescriptions for reducing the primordial quadrupole may have problems
producing \cte amplitudes as high as those that are observed on the largest
angular scales, given the correlation between \ctt and \cte.

Most of the other multipole bins up to $\ell<20$ in Figure
\ref{fig:wmap1yr_all} appear reasonably consistent with the best fit
cosmological model. The $\ell=20-23$ bin is discrepant at nearly the same
level in the \ctt power spectrum, but at current noise levels the degree of
correlation between the measured \cte and \ctt is negligible. The full
four-year data of WMAP will help to improve these constraints. More
accurate treatments of the errors and residual correlation due to the
cut-sky will be possible. Furthermore, it would be particularly interesting
to study the value of the probability ${\cal P} (\C_\l^{TE}|\C_{\l
0}^{TE})$ for various galactic cuts as a probe of the galactic contribution
(D. Spergel, private communication). In the more distant future, data of
the quality forecasted for the Planck
satellite\footnote{http://www.astro.esa.int/SA-general/Projects/Planck/}
will provide even more powerful consistency checks on the best-fit
cosmological model. More accurate modeling of the galactic emission will
also certainly help to address those issues.

{\it What would it mean if the observed statistics of the CMB do not appear
to be consistent with the best fit cosmological model?} The simplest
explanation would be that at least one of the measured components is not
purely cosmological in origin, possibly due to Galactic contamination.  For
example, removing the Galactic foreground appears to enhance the inferred
quadrupole \citep{deOl03}.  The absence of foreground contamination would
be an exciting indication of new physics.

\acknowledgments { This work was supported by NASA ATP grants NAG5-7154,
NAG5-13292, NSF grant AST-0204514, and the W.M. Keck Foundation. We have
benefited greatly from discussions with Carlo Contaldi, Eiichiro Komatsu, 
Lloyd Knox, Hiranya Peiris, David Spergel, Licia Verde and Matias
Zaldarriaga.  A.L. gratefully acknowledges sabbatical support from the
John Simon Guggenheim Memorial Fellowship and the Institute for
Advanced Study.}


\begin{thebibliography}{}
\vspace{0.5cm}

\bibitem[Bennett \etal (2003)]{Be03} Bennett, C.L., \apjs, 148, 1 
\bibitem[Bennett \etal (1996)]{Be96} Bennett, C.L., \apj, 464, L1
\bibitem[Bond (1995)]{Bo95} Bond, J.~R.\ 1995, Physical Review Letters, 74, 4369
\bibitem[Bridle \etal (2003)]{Bri03} Bridle, A., Lewis, A.M.,
  Weller, J., Efstathiou, G. 2003, NewAr, 47, 8, 787 
\bibitem[Contaldi \etal (2003)]{Con03} Contaldi, C.R., Peloso, M., 
Kofman, L., Linde, A.  2003, JCAP, 07, 002 
\bibitem[Cline, Crotty, \& Lesgourgues(2003)]{cline03} Cline, J.~M., 
Crotty, P., \& Lesgourgues, J. 2003, JCAP, 09, 010
\bibitem[de Oliveira-Costa \etal (2004)]{deOl03}  de Oliveira-Costa, A., 
Tegmark, M., Zaldarriaga, M. \& Hamilton, A. 2004, \prd, 69, 6, 063516
\bibitem[Efstathiou (2003a)]{Ef03a} Efstathiou, G. 2003, \MNRAS, 343, 4, L95 
\bibitem[Efstathiou (2003b)]{Ef03b} Efstathiou, G. 2003, \MNRAS, 346, 2, L26
\bibitem[Feng and Zhang (2003)]{Fe03} Feng, B., and Zhang, X. 2003,
Physics Letters B, 570, 3-4, 145
\bibitem[Hinshaw \etal (2003)]{Hi03} Hinshaw, G. \etal, \apjs, 148, 135
\bibitem[Kaplinghat \etal (2003)]{Ka03} Kaplinghat, M., \etal, 2003, \apj, 
583, 24 
\bibitem[Hu and Dodelson(2002)]{hu02} Hu, W., and Dodelson, S. 2002, \araa, 40, 171
\bibitem[Jaffe(2003)]{jaffe03} Jaffe, A.~H. 2003, NewAr, 47,
11, 1001 
\bibitem[Kogut \etal (2003)]{Ko03} Kogut, A. \etal, \apjs, 148, 161
\bibitem[Leitch et al.(2002)]{Le02} Leitch, E.~M.~\etal 2002, \nat, 420, 763 
\bibitem[Spergel \etal (2003)]{Sp03} Spergel, D. \etal 2003, \apjs, 148, 175
\bibitem[Tegmark \& Zaldarriaga(2002)]{TeZa02} Tegmark, M.~\& Zaldarriaga, M.\ 2002, \prd, 66, 103508  
\bibitem[Tegmark \etal (2003)]{Te03} Tegmark, M., de Oliveira-Costa, A. 
\& Hamilton, A. 2003, \prd, 68, 12, 123523  
\bibitem[Verde \etal (2003)]{Ve03} Verde, L. \etal, \apjs, 148, 195; see also
  \texttt{http://lambda.gsfc.nasa.gov} for relevant data and routines
\bibitem[Zaldarriaga(1997)]{Za97} Zaldarriaga, M.\ 1997, \prd, 55, 1822 
\bibitem[Zaldarriaga \& Seljak (1997)]{ZaSe97} Zaldarriaga, M. \& Seljak, U. \ 1997, \prd, 55, 1830 
\bibitem[Zaldarriaga, Spergel, \& Seljak(1997)]{ZaSpSe97} 
Zaldarriaga, M., Spergel, D.~N., \& Seljak, U.\ 1997, \apj, 488, 1 

\end{thebibliography}
\end{document}